\documentclass[aps,prb,twocolumn,groupedaddress,floats,showpacs]{revtex4}

\usepackage{latexsym}
\usepackage{dcolumn}
\usepackage[dvips]{graphicx}
\usepackage{amssymb,amsmath,bm}
\usepackage{graphics}
\usepackage{epsf}

\begin{document}

\draft

\title{Transport and drag in undoped  electron-hole bilayers}

\author{E. H. Hwang and S. Das Sarma}
\address{Condensed Matter Theory Center, 
Department of Physics, University of Maryland, College Park,
Maryland  20742-4111 } 

\date{\today}

\begin{abstract}

We investigate transport and Coulomb drag properties of
semiconductor-based  
electron-hole bilayer systems.
Our calculations are motivated by recent experiments in
undoped electron-hole bilayer structures based on GaAs-AlGaAs gated
double quantum well systems.  
Our results indicate that the background charged impurity
scattering is the most dominant resistive scattering mechanism in the
high-mobility bilyers.
We also find that the drag transresistivity is 
significantly enhanced when the electron-hole layer separation is small
due to the exchange induced renormalization of the single
layer compressibility.


\end{abstract}

\maketitle

\vspace{0.5cm}


\section{introduction}

Much interest has recently focused on bilayer semiconductor
structures\cite{drag,Eisenstein,bilayer}, where two quantum wells are
put in close proximity with 
a high insulating barrier between them to suppress interlayer
tunneling. Such systems are of intrinsic interest because competition
between intralayer and interlayer Coulomb interaction (i.e., correlation
effects), particularly at low carrier densities when Coulomb
interaction dominates (or in high magnetic fields, when the
electronic kinetic energy is quenched), may lead to exotic ground
states, collective properties, and quantum phase transitions. Although
much of the recent work has concentrated on bilayer quantum Hall
systems \cite{Eisenstein}, the essential issue of the competition
between electron 
kinetic energy and intra-/interlayer Coulomb correlations is quite
generic, and interaction-induced novel ground states and  collective
properties are possible even in the zero field
situation if the carrier density is low enough and the two layers are
close enough to produce strong inter-layer interaction.

Significant recent progress \cite{bilayer} has been made in the
fabrication of two dimensional (2D) 
electron-hole bilayers because of intense interest in exciton
condensation effects that are predicted 
to occur  at low carrier densities
and close proximity of the two-dimensional electron gas (2DEG) to the
two-dimensional hole gas (2DHG) \cite{ehp}. 
At sufficiently low densities the attractive interlayer
Coulomb interaction between an electron in one layer with a hole in
the other layer would dominate over the intralayer repulsive
interaction. An excitonic phase with electron-hole pairs (EHP) may
form under such conditions, being lower in energy than the independent
electron and hole Fermi liquid layers. Unlike excitons in bulk
semiconductors, these EHPs would not annihilate by emitting radiation
because the electrons and holes are spatially separated in two 
different layers.

Exciton condensation has previously been reported in
electron or hole, but not electron-hole, bilayer quantum Hall systems
\cite{Eisenstein}, but 
at a zero magnetic field 
the measurement of exciton effects in electron-hole (e-h) bilayers 
has proven to be extremely difficult because of the 
limitation of adjusting the densities in the two layers, in the low
density regime, to 
match up the density of the 2DEG to the density of the 2DHG
\cite{bilayer}.  However, recent
fabrication of e-h bilayer systems \cite{Seamons,Seamons2} 
based on the heterostructure 
insulated-gate field-effect transistors (HIGFETs) \cite{Kane,lilly} 
allows for independent contacts to each layer as well as high
mobility and  independently tunable low densities for the 2DEG and 2DHG. 
Recent experiments  \cite{Seamons2} in very low density and high
quality e-h bilayers on HIGFETs
show anomalous drag resistance behavior. The e-h
bilayers on HIGFETs are undoped gated double quantum well devices, so
the transport properties of the device may be 
different from the previously investigated modulation doped
e-h bilayer systems \cite{bilayer}. 
Recent experiments \cite{Seamons} have reported on the
density dependent 2D mobilities of the electrons and holes in e-h
bilayer in the same low density regime where the anomalous drag
resistivity has been observed \cite{Seamons2}.

Our goal in this paper is to develop detailed and realistic
quantitative theories for density-dependent transport and drag in 2D
e-h bilayers using the Fermi liquid picture for both carrier
systems. For transport calculations, we consider Coulomb scattering by
unintentional random charged impurities invariably present in the
background. For drag calculations, we consider the Coulomb interaction
between the electron and the hole layers. The important ingredient of
physics included in both theoretical calculations is wave vector and
frequency dependent dielectric screening by the electrons and holes
themselves  --- for the charged impurity scattering induced transport
calculation, the relevant screening is static since the impurities are
quenched in the background whereas for the e-h frictional drag, the
full dynamical screening must be considered in the theory. Our aim
(and hope) in this work is to assist in the experimental detection of
the excitonic phase in e-h bilayers through a careful comparison
between our theory and the recent measurements \cite{Seamons,Seamons2}
where any qualitative (or even dramatic quantitative) difference
between our theory and the experimental data could provide suggestive
evidence (or at least, interesting clues) for the failure of the Fermi
liquid picture in the physical system, thereby indicating the possible
emergence of a novel non-Fermi liquid collective phase.

In this paper, motivated by recent experiments on low-density e-h
bilayers \cite{Seamons,Seamons2}, we study the transport properties of
undoped electron-hole bilayers fabricated on HIGFETs. We
calculate both the mobilities of each layer and the frictional Coulomb
drag resistivity.  
In order to understand the low temperature 2D transport properties
we have carried out a microscopic transport calculation using the
Boltzmann theory \cite{dassarma1,RMP}.  
Since the HIGFET is an undoped heterostructure,
the 2D carrier resistivity is limited by screened
background random Coulomb scattering in high mobility systems. 
With the unintentional background random 3D charged impurities as the
main scattering center we obtain
good agreement with the recent transport experiments \cite{Seamons} in
undoped e-h bilayer systems.

In order to calculate the frictional drag resistivity of undoped
electron-hole bilayers we use
the many body-Fermi liquid  
diagrammatic perturbation theory with dynamically screened
electron-electron interaction \cite{drag,hwang1}.
The recent experiments in electron-hole bilayers with 
small layer separations and
especially at low temperatures and densities
\cite{Seamons2} show that the measured Coulomb
drag increases as the temperature decreases at the lowest measured
temperatures, which can not be  explained based on the Fermi
liquid theory, since the Fermi liquid theory predicts the
low-temperature $T^2$ (or $T^2\ln T$) behavior of drag resistivity
\cite{drag}. This anomalous behavior in the e-h drag may be related to
the exciton condensation. 
However, our calculation shows good agreement with the
experimentally observed drag data for the electron-hole bilayer with
large separation, and even bilayers with small separation
above the critical temperature of possible exciton condensation.
The reasonable agreement between our theory and the measured drag
\cite{Seamons2} in the larger layer separation regime for all
temperatures and the smaller separation regime for higher
temperatures is in some sense tantalizing circumstantial evidence in
support of the emergence of the collective excitonic phase in the
measurements reported in ref. \onlinecite{Seamons2}, where the drag
shows a minimum as a function of temperature and increases with
decreasing temperature at the lowest temperature.

We describe our background charged impurity limited transport theory
and results in Sec. II, followed by our e-h Coulomb drag results in
Sec. III. We conclude in Sec. IV with a discussion.

\section{transport}

We start by writing down the theoretical formulae for
conductivity $\sigma$ in the many-body Fermi liquid 
RPA-Boltzmann theory approximation widely used in the literature
\cite{dassarma1,RMP}.
The carrier conductivity $\sigma$ is given by 
$\sigma = ne \mu$, 
where $n$ is the carrier density and the carrier mobility $\mu$ is
given by $\mu = e \langle \tau \rangle/m$. Here $m$ is the carrier
effective mass  and  $\langle \tau(E) \rangle$ is the energy averaged
transport relaxation time.
We calculate the impurity ensemble averaged relaxation time $\tau(E)$
due to elastic disorder scattering by the background quenched charged
impurities in the Born approximation: 
\begin{eqnarray}
\frac{1}{\tau(E_k)} & = & \frac{2\pi}{\hbar}\sum_{\alpha}
\int \frac{d^2k'}{(2\pi)^2}  
\int^{\infty}_{-\infty}dzN_i^{(\alpha)}(z) \nonumber \\
& \times & |u^{(\alpha)}({\bf k}-{\bf k}';z)|^2 
(1-\cos \theta_{{\bf k k}'})
\delta(E_k-E_{k'}), 
\label{itau}
\end{eqnarray}
where $E(k) = \hbar^2 k^2/2m$ is the 2D carrier energy for 2D wave
vector {\bf k}; $\theta_{{\bf k k}'}$ is the scattering angle between
carrier scattering wave vectors {\bf k} and ${\bf k}'$; the delta
function $\delta(E_k-E_{k'})$ assures energy conservation for
elastic scattering due to charged impurities where the screened
scattering potential is denoted by $u^{(\alpha)}({\bf q};z)$ 
with ${\bf q} \equiv
{\bf k} - {\bf k}'$ as the 2D scattering wave vector and $z$ is the
quantization or the confinement direction normal to the 2D layer. The
quantity $N_i^{(\alpha)}(z)$ in Eq. (\ref{itau}) denotes the 3D charged
impurity density (with the $z$ dependence reflecting a possible
impurity distribution) of the $\alpha$-th kind with $\alpha$
representing the various possible types of impurities which may be
present in 2D semiconductor structures. However,
we  emphasize
that there is no intentional doping in the HIGFETs used in
refs. \onlinecite{Seamons, Seamons2}. Therefore, we use
the unintentional random background 3D impurities as the only scattering
source in our calculation
to keep the number of unknown parameters a minimum -- this background
random 3D impurity
density essentially sets the overall scale of resistivity in our
results. We emphasize that we can obtain good qualitative agreement
with experimental data by choosing three different 
kinds of charged impurities (i.e. interface, remote, and bulk)
parameterized by a few reasonable parameters, but we do not see much
point in this data fitting-type endeavor, and therefore keep only one
unknown parameter in the theory, $n_{3D}$, which is the 3D density of
the unintentional charged impurities, which are assumed to be
uniformly and randomly distributed the background throughout the whole
sample. As emphasized above, this only lets the scale of the overall
resistivity, not its qualitative behavior.

In Eq. (\ref{itau}) the screened impurity potential
$u^{(\alpha)}(q;z)$ is given by 
$ u^{(\alpha)}(q;z)\equiv V^{(\alpha)}_{\rm{imp}}(q;z)/\epsilon(q) $
where $V^{(\alpha)}_{\rm imp}$ is the bare potential due to a charged
impurity and 
$\epsilon(q)$ is the carrier dielectric screening function which is
necessary since the charged impurity potential is Coulombic.
The bare impurity potential is given by 
\begin{equation}
V^{(\alpha)}_{\rm imp}(q;z) = \frac{2\pi Z^{(\alpha)}e^2}{\kappa q}
F^{(\alpha)}_{\rm imp}(q;z), 
\end{equation}
where $Z^{(\alpha)}$ is the impurity charge strength, $\kappa$ is the
background (static) lattice dielectric constant, and $F_{\rm imp}$ is a
form factor determined by the location of the impurity and the subband
wavefunction $\psi(z)$ defining the 2D quantum well confinement. 
The finite wave vector dielectric screening function is written in the
RPA as  
\begin{equation}
\epsilon(q) = 1- v(q) \Pi(q,T),
\end{equation}
where $\Pi(q,T)$ is the 2D irreducible finite-temperature (and finite
wave vector)  polarizability function 
and $v(q)= v_{2D}(q) f(q)$
is the effective bare electron-electron (Coulomb) interaction in the
system with $v_{2D}(q) = 2\pi e^2/(\kappa q)$ being the 2D Fourier
transform of the  Coulomb potential, and
$f(q)$ being the Coulomb form factor arising from the subband
wavefunctions $\psi(z)$.

In Fig. 1 the sample configuration, following the experimental
references \cite{Seamons,Seamons2}, used in our calculation is shown
schematically. The electron and hole quantum wells with widths $a=200$
\AA \; are separated by the $d=300$ \AA \; barrier. Since there is no
intentional doping, the charged Coulomb impurities are distributed randomly
with a constant 3D impurity density $n_{3D}$ throughout the sample (both
barriers and quantum wells). Our sample
configuration  resembles the experimental setup of
refs. \onlinecite{Seamons,Seamons2}. 

\begin{figure}
\epsfysize=2.5in
\centerline{\epsffile{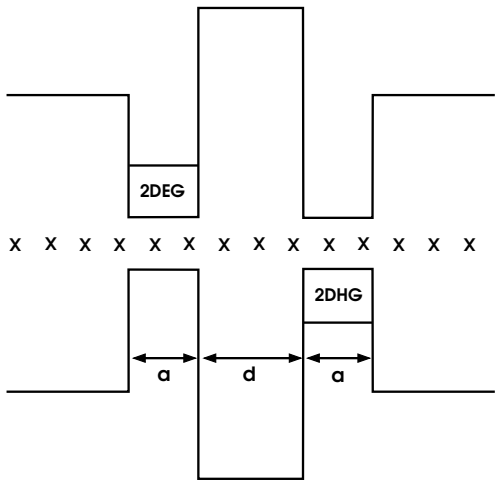}}
\caption{The sample configuration used in this calculation is shown
  schematically. In this 
  calculation we use the parameters $a=200$ \AA, $d=300$ \AA. ``x''
  denotes the charged impurities which are distributed uniformly in
  the background with a 3D effective density of $n_{3D}$.}
\end{figure}

In Fig. 2 we show (a) the calculated density dependent mobility of each
layer and (b) the ratio of hole mobility to electron mobility at fixed
temperature $T=300$ mK for the sample configuration 
given in Fig. 1. The symbols indicate some representative experimental
data points taken from ref. \onlinecite{Seamons}. 
We use a fixed charged impurity density $n_{3D} = 4.2 \times
10^{14}$ cm$^{-3}$ and the known effective mass $m=0.067m_e$ for
electrons and $m=0.3m_e$ for holes, respectively. 
As emphasized above the only resistive carrier scattering included in
the results of Fig. 2 is that by background charged impurities. (We
discuss later the issue of inter-layer e-h scattering itself
contributing to 2D transport, arguing it to be negligibly small).
We assume a single value of $n_{3D}$ for calculating electron and hole
mobilities in the different layers. We include the effects of
confinement potential through the infinite square well confinement
model, which should be excellent for the samples of
refs. \onlinecite{Seamons} and \onlinecite{Seamons2}.
We neglect all phonon scattering effects because
our theoretical estimate shows phonon scattering to be negligible for
2D carriers in GaAs structures in the $T<1K$ regime of interest to us.
The phonon scattering becomes relevant 
typically in the $T > 3K$ regime \cite{dassarma2}.

\begin{figure}
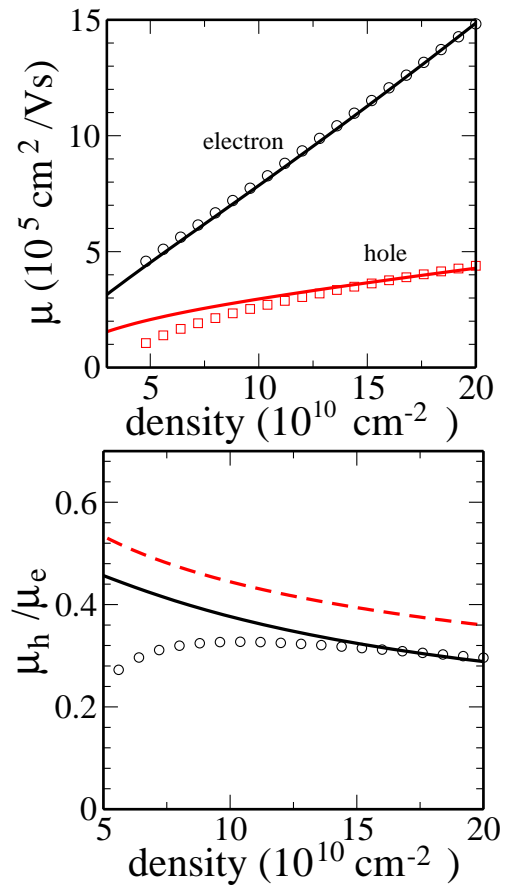

\epsfysize=2.3in
\centerline{\epsffile{fig2a.eps}}
\epsfysize=2.3in
\centerline{\epsffile{fig2bb.eps}}
\caption{(Color online)
(a) Calculated mobilities of electron ($m=0.067$) and hole ($m=0.3$)
as a function of carrier density. The symbols are experimental data
taken from ref. [5].
(b) The ratio of hole mobility to electron mobility. The solid lines
are full theory and the dashed line in (b) is the simple approximation given
in the Eq. (4) of the text. Here we use $n_{3D} = 4.2 \times 10^{14}cm^{-3}$.
}
\end{figure}

For the experimental density range ($n \sim 5\times 10^{10}cm^{-2} - 2
\times 10^{11} cm^{-2}$) our theoretically calculated density
dependent electron mobility  
has an effective exponent $\alpha \approx 1$ in $\mu \propto n^{\alpha}$,
which is consistent with the experimental data of
ref. \onlinecite{Seamons}, indicating 
that the screened background charged impurity is the main scattering
source in the sample.
For the density dependent hole mobility we can not explain the
experimental data for all densities 
by a single power law behavior.
At high hole densities, $p>10^{11}cm^{-2}$
the calculated mobility approaches an approximate power law
behavior  $\alpha \approx 0.7$, which is in excellent agreement with
the experimental data of ref. \onlinecite{Seamons}.  At low hole
densities, $p <   
10^{11} cm^{-2}$, the power law changes to a much stronger function of
density, pointing to a change in the screening properties of the
system at low carrier density. 
Although our theoretical result in Fig. 2(a) indicates a changing
exponent $\alpha$ for the holes at lower ($<10^{11}cm^{-2}$) hole
densities, the experimental $\alpha$ is much more strongly affected by
the decreasing density than our RPA Boltzmann transport theory.
We emphasize that the
discrepancy between the low density experimental data 
and our calculation indicates that the screening theory,
which captures the essential experimental features at higher densities,
does not describe the hole transport very well at the lower range of
the experimental density.

There are several understandable reasons for the systematic discrepancy
between our theory and the lower density hole data. First, the holes,
being more massive ($m_h = 0.3m$ versus $m_e=0.07m$), have stronger
interaction effects than electrons as characterized by the
dimensionless interaction coupling strength $r_s = (\pi n)^{-1/2}/a_B$
where $a_B=\kappa \hbar^2/me^2$ is the effective Bohr radius, which for
the holes varies between $r_s \approx 12-6$ ($n_h = 5\times 10^{10} -
2\times 10^{11} cm^{-2}$) compared with $r_s \approx 2.7-1.3$ for the
electrons. The large value of the hole interaction parameter $r_s \sim 12
$ at the hole density of $5\times 10^{10} cm^{-2}$ indicates that our
RPA-Boltzmann theory may be less quantitatively reliable at lower hole
densities than at lower electron densities since the theory is really
quantitatively valid only at small $r_s$, where, of course, we get
excellent agreement with the experimental data both for electrons and
holes. Second, and perhaps more importantly, charged impurities
introduce strong density inhomogeneity at low carrier densities as
linear screening eventually breaks down leading to a 2D metal to 2D
insulator transition (2D MIT) \cite{Dassarma,Manfra}, thereby causing
a systematic failure of 
our linear screening RPA-Boltzmann theory at low enough densities. While such
a density inhomogeneity driven failure of our linear RPA-Boltzmann
theory is inevitable at low enough carrier densities for both the
electrons and the holes, the critical density for the hole 2D MIT
\cite{Manfra} is
typically much larger than the corresponding electron critical
density \cite{Dassarma}. We thus expect that our RPA-Boltzmann theory
will deviate 
more from the experimental data for 2D holes than for 2D electrons at
lower carrier densities, particularly since the 2D holes have lower
mobilities than the 2D electrons. We believe that this nonlinear
screening effect along with the large $r_s$ values are responsible for
the discrepancy between theory and experiment for the low density
hole results shown in Fig. 2.

In Fig. 2(b) we show the ratio of hole mobility to electron mobility
as a function of carrier density. In the high density regime, $n>10^{11}
cm^{-2}$, the measured experimental mobility ratio is about 0.3 and
decreases gradually as density increases. Our calculated $\mu_h/\mu_e$
shows excellent
agreement with experimental data in the high density limit, but
deviates from the experiment at low ($<10^{11}cm^{-2}$) density for
reasons discussed above (i.e. large $r_s$ and density inhomogeneity). 
For screened Coulomb scattering, assuming strong screening, we can
easily derive the approximate qualitative formula for the mobility
by ignoring the wave vector dependence in Eq. (1) for transport:
\begin{equation}
\mu \propto (2k_F +q_{TF})^2/m^2,
\end{equation}
where $k_F=(2\pi n)^{1/2}$ is the Fermi wave vector and $q_{TF}=2/a_B$
is the Thomas Fermi screening wave vector.
Thus, in the high density limit ($q_{TF}/2k_F \ll 1$, typically $n
>10^{13}cm^{-2}$ for GaAs systems) 
the mobility is proportional to the inverse square of the effective
mass, (i.e. $\mu \propto 1/m^2$).
However, in the low density limit (or $q_{TF}/2k_F \gg 1$)
the mobility is independent of the carrier mass.
Thus, in the simple theory the ratio of mobility increases to unity as
density 
decreases. But the measured mobility ratio decreases as density decreases
below $n < 10^{11}cm^{-2}$. Again this experimental behavior arises
from nonlinear screening and the large $r_s$ values at low carrier
density.

In Fig. 2(b), in addition to the full numerically calculated
RPA-Boltzmann transport results and the experimental results from
ref. \onlinecite{Seamons}, we also show the simple results for
$\mu_h/\mu_e$ derived on 
the basis of the qualitative analytic formula of Eq. (4). It is quite
interesting to note that in the experimental density range covered in
the measurements of ref. \onlinecite{Seamons}, the measured
$\mu_h/\mu_e \approx 
0.28-0.32$ is almost a density independent constant, which does not
differ much from the simple electron-hole effective mass ratio
$\mu_h/\mu_e \approx m_e/m_h \approx 0.23$ which is very different
from either the high density ($q_{TF} \ll 2k_F$) asymptotic value
$\mu_h/\mu_e = (m_e/m_h)^2 \approx 0.05$ or the low density ($q_{TF}
\gg 2k_F$) asymptotic value $\mu_h/\mu_e \approx 1$. The disagreement
of the approximate formula, Eq. (4), with the experimental data can be
easily understood by the realizing that the dimensionless parameter
$q_{TF}/2k_F =r_s/\sqrt{2}$ changes between 0.92 and 1.9 for electrons and
4.2 and 8.5 for holes in the experimental density range of Fig. 2, and
therefore neither the strong screening nor the weak screening
approximation applies. It is, however, curious that the actual
experimental (and the full theoretical) values of $\mu_h/\mu_e \sim
m_e/m_h$ in the experimental density range studied in Fig. 2, which
(i.e. $\mu \propto \sigma \propto m^{-1}$) would follow trivially from
the Drude conductivity formula $\sigma =ne^2 \tau/m$ if one assumes
$\tau$ to be a constant independent of effective mass as is often done
in the literature. This apparent approximate agreement of the trivial
$\mu_h/\mu_e \sim m_e/m_h$ relationship with the experimental results
of ref. \onlinecite{Seamons} is, however, purely fortuitous, and
should not be taken 
seriously. For very high (low) electron and hole density, there is no
doubt that $\mu_h/\mu_e \sim m_e^2/m_h^2$ ($\sim 1$), but both the
asymptotic high and low density regimes would be difficult to achieve
experimentally, and in smoothly going between these two regimes, the
intermediate density regime of experimental interest only seems to
obey the simple $\mu_h/\mu_e \sim m_e/m_h$ relation since this
$m_e/m_h$ behavior is obviously intermediate between $(m_e/m_h)^2$ and
$(m_e/m_h)^0=1$.

It is, however, interesting to note that the approximate mobility
formula defined by Eq. (4), where the charged impurity scattering
strength is characterized by the constant momentum transfer of
$(2k_F+q_{TF})$ completely ignoring the wave vector dependence of
Coulomb interaction and assuming Thomas-Fermi screening so that the
$1/q$ Coulomb scattering strength is parameterized simply by
$1/(2k_F+q_{TF})$ corresponding to the $q=2k_F$ backward carrier
scattering across the 2D Fermi surface, describes well the density
dependence of mobility as compared with the full RPA-Boltzmann
theory. Both results disagree with experiments at low carrier
densities for reasons discussed above.

\section{coulomb drag}

The principal motivation \cite{Seamons,Seamons2} behind trying to
fabricate closely spaced 
e-h bilayers with small interlayer separation is to probe the
interlayer e-h Coulomb 
interaction. The bilayer frictional drag, which is a direct probe of
interlayer correlations, in a many
body-Fermi liquid  
diagrammatic perturbation theory with dynamically screened
electron-electron interaction is given by \cite{drag,bilayer}
\begin{equation}
\rho_D=\frac{\hbar^2}{2 \pi e^2np k_BT}\int\frac{q^2 d^2q}{(2\pi)^2}
\int{d\omega} \frac{F_e(q,\omega)F_h(q,\omega)}
{\sinh^2(\beta \omega/2)},
\label{rho_drag}
\end{equation}
where $F_{e,h}(q,\omega) = |u_{eh}^{sc}(q,\omega)| \rm{Im}
\Pi_{e,h}(q,\omega)$, with $u_{eh}^{sc} = v_{eh}^c
/\epsilon(q,\omega)$ is the dynamically screened interlayer Coulomb
interaction between the electron and the hole layers, and
$\Pi(q,\omega)$ is the 2D 
polarizability. 
Note that the dielectric
function ${\epsilon}(q,\omega)$ entering
Eq. (\ref{rho_drag}) is a 2-component tensor and is given by
\cite{dassarma3} 
\begin{eqnarray}
|\epsilon(q,\omega)|& = &\left [1-v_{e}(q)\Pi_{e}(q,\omega) \right ]
\left [1-v_{h}(q)\Pi_{h}(q,\omega) \right ] \nonumber \\
& - & v_{eh}(q)v_{he}(q)\Pi_{e}(q,\omega)\Pi_{h}(q,\omega),
\end{eqnarray}
where $e,h$ correspond to electron and hole layers.
(For details on the drag formula and its implications, see
refs. \onlinecite{drag,bilayer}.) With the assumption of a large
inter-layer separation $d$ ($k_{F}d\gg 1,$ $q_{TF}d\gg 1$)
it is easy to show that within RPA, where vertex corrections are
neglected in the dynamical polarizability $\Pi$, the drag resistivity
is given at low temperatures ($T \ll T_F$) by the simple formula:
\begin{equation}
\rho_{D}=\frac{m_e m_h}{n p e^2}
\frac{\zeta (3)}{16 \hbar}
  \frac{(k_{B}T)^2}{k_{F_e}k_{F_h}q_{TF_e}q_{TF_h}d^{4}} 
\label{rhod}
\end{equation}
with $\zeta$ being the Riemann zeta function. 
This result shows that $\rho_D(T,d) \propto T^2/d^4$.
However, we note that Eq. (\ref{rhod}) is valid 
only for high density and low temperature  ($T/T_F \ll 1$) as well
as large separation. At low
densities or small layer separations (i.e. $k_Fd \alt 1$) the actual
drag is much enhanced compared with the simple 
formula of Eq. (7) as has earlier been discussed \cite{hwang1}.

\begin{figure}
\epsfysize=2.4in
\centerline{\epsffile{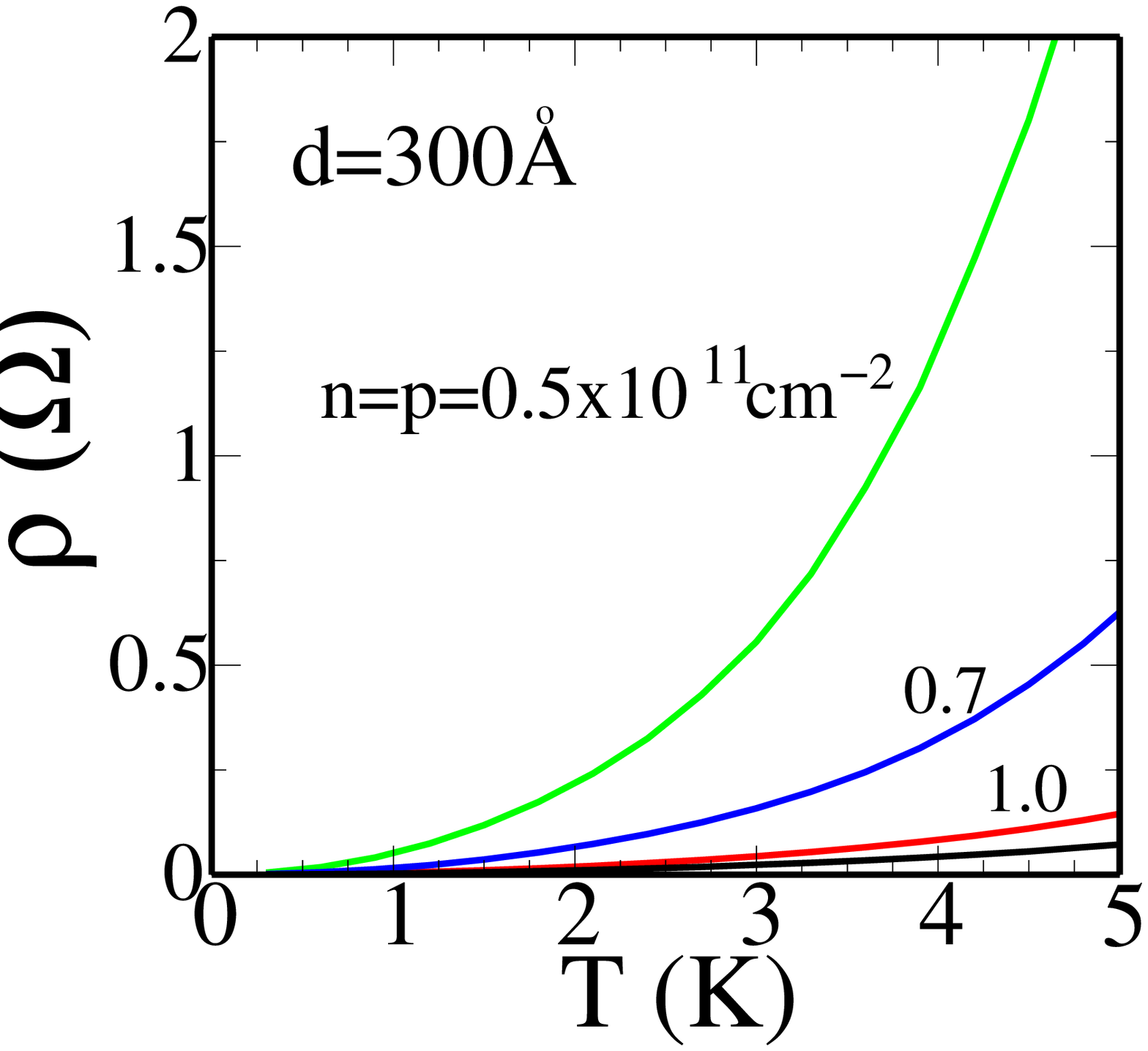}}
\epsfysize=2.4in
\centerline{\epsffile{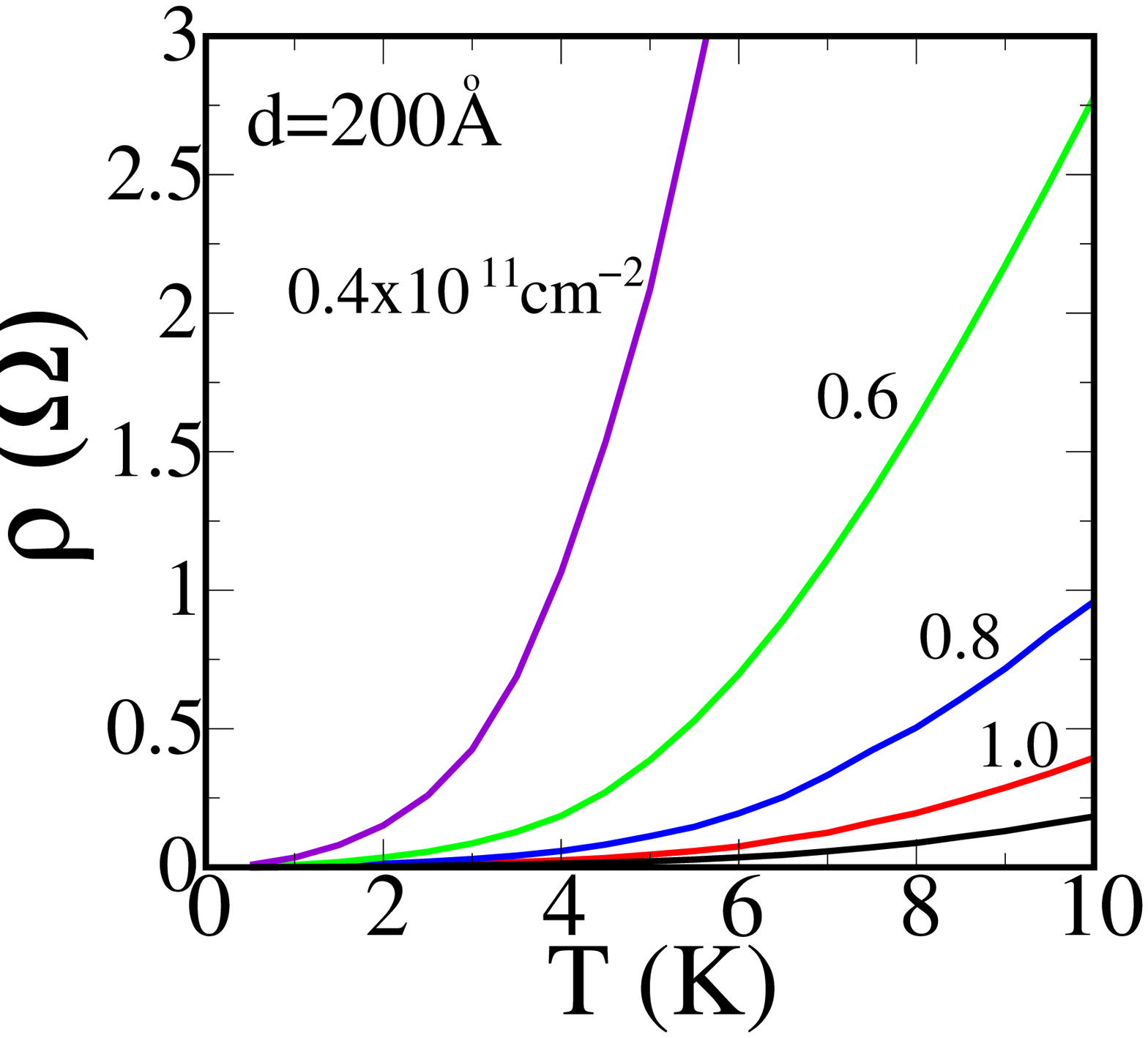}}
\caption{(Color online)
Calculated Coulomb drag of electron-hole bilayer systems with 
(a) $d=300$ \AA \; barrier and (b) $d=200$ \AA \; barrier. 
In (a) we use the matched electron-hole density of 
$n=p=$5, 7,  10, 12 $\times
10^{10}cm^{-2}$ and in (b) $n=p=$4, 6, 8, 10, 12 $\times
10^{10}cm^{-2}$.
}
\end{figure}

In our calculation we include the
finite layer thickness effect and relax the 
condition $k_{F}d \gg 1$
because  $k_{F}$ and $d$ are small in the samples 
of ref. \onlinecite{Seamons2} 
-- this makes large momentum scattering important in 
contrast to the usual $k_{F}d \gg 1$ case where carrier backscattering
is neglected.
In Coulomb drag between two high density 2D electron layers with
large layer separation (i.e. $k_Fd \gg 1$)
the backward scattering $q \approx 2k_F$ is ignored
due to the exponential dependence of the interlayer Coulomb
interaction $V(q) \propto \exp(-qd)/q$ on $d$. In this case the
Coulomb drag is
dominated by small angle scattering and 
Eq. (\ref{rhod}) is a very good approximation.
In the e-h bilayer of small $d$ (and low density) in
ref. \onlinecite{Seamons2}, large 
angle scattering becomes important and Eq. (7) would not apply at all,
and one needs to carry out the full calculation of Eqs. (5) and (6) to
obtain the e-h drag.

We directly theoretically calculate the e-h
Coulomb drag using Eqs. (5) and (6) without making any additional
assumptions and using the sample parameters of
refs. \onlinecite{Seamons} and \onlinecite{Seamons2}. 
In  Fig. 3 we show the calculated drag resistivity of electron-hole
bilayer with each well width of $a=200${\AA} as a function of temperature 
for several matched e-h densities. Here the electron 
layer is the driving layer and the hole layer is the drag layer,
i.e. the current $I_e$ is applied in the electron layer and drag voltage
$V_D$ measured in the hole layer with $\rho_D = V_D/I_e$. 
In Fig. 3(a) we use the density $n=p=$5, 7,  10, 12 $\times
10^{10}cm^{-2}$ and a layer separation of $d=300$ \AA. 
In the experiments of ref. \onlinecite{Seamons2} the observed low density 
drag resistivity ($n=p=5\times 10^{10}cm^{-2}$) is
orders of magnitude larger than the corresponding theoretical drag result 
of Eq. (\ref{rhod}), 
but our calculation  with Eq. (\ref{rho_drag}) shows 
good agreement with the experimental data 
within a factor of 2.
In e-h bilayers the low-energy acoustic plasmon lies inside the
single particle excitation region of the heavier carriers
\cite{dassarma4}, which gives 
rise to a tremendous enhancement of drag resistivity not captured at
all in Eq. (7).
We include this effect by 
going beyond the simple 
RPA (valid at high density) in calculating the polarizability by
incorporating vertex corrections through
local field corrections.
The strong correlation effect in the low densities gives to a
large enhancement of the drag  due to the presence of coupled 
plasmon modes.

In Fig. 3(b) we use the density $n=p=$4, 6, 8, 10, 12 $\times
10^{10}cm^{-2}$ and a  layer separation of $d=200$ \AA.
With this narrow barrier sample, Seamons {\it et al.} \cite{Seamons2}
find very interesting and unexpected drag resistivity which is not
observed in the $d=300$ \AA \; barrier sample of
ref. \onlinecite{Seamons2}. 
Even though the measured drag resistivity show $\rho_D \propto T^2$
behavior above $T \approx 1K$,
for temperatures below 1K an upturn of the Coulomb drag 
is measured with decreasing temperatures (instead of $\rho_D
\rightarrow 0$ as $T^2$, 
$\rho_D$ increases as $T\rightarrow 0$). The characteristic temperature
below which the observed $\rho_D$ 
increases anomalously in ref. \onlinecite{Seamons2} with decreasing
temperature itself increases 
with density. Thus, the
upturn in drag at low 
temperature can not be explained within our Fermi liquid theory and
may signal the formation of a novel phase in electron-hole 
system. However our calculation at temperatures above $T=1K$ shows
good  agreement with the experimental data and we also get good
agreement at all temperatures for sample with larger $d$, neither of
which would happen with the simple formula of Eq. (7) which disagrees
with experiments by orders of magnitude.

\section{discussion and conclusion}

Before conclusion, we return to transport and discuss the layer
independence of transport in the 
electron-hole bilayer. 
This layer independence is reflected in the experimental observation
\cite{Seamons, Seamons2} of the mobility in one layer (e.g. the
electron layer) to be independent of the carrier density in the {\it
  other} layer (i.e. the hole layer) and vice versa.
In the presence of the adjacent layer
the total scattering time of one layer can be expressed as
${1}/{\tau_t} = {1}/{\tau_i} + {1}/{\tau_D}$, where $\tau_i$ is the
scattering time due to the background charged impurities given in
Eq. (\ref{itau}) and shown in Fig.2,  and
$\tau_{D}$ is  
the e-h scattering time due to the inter-carrier inter-layer
scattering between electrons and holes  which is
given by $\tau_D = m/ne^2\rho_D$, with $\rho_D$ being our calculated
inter-layer e-h frictional drag resistivity [Eq. (5) and
Fig. 3]. However, for samples we 
consider in this paper the typical scattering times are $\tau_i \sim
ns$ and 
$\tau_D \sim \mu s$ below $T = 2K$. i.e. $\tau_i^{-1} \gg
\tau_D^{-1}$. Therefore we expect the mobility 
of each layer to be dominated by impurity scattering and
entirely independent of the adjacent layer density.
The extremely small measured and calculated values of the interlayer
Coulomb drag $\rho_D$, corresponding to extremely long drag relaxation times
of microseconds, imply that interlayer e-h scattering makes a very
small ($\sim 0.1 \%$) contribution to the measured $dc$ conductivity
(i.e. 2D mobility) of each layer. The measured 2D mobility in the
experimental e-h bilayers of refs. \onlinecite{Seamons} and
\onlinecite{Seamons2} is therefore entirely
dominated by the background charged impurity scattering. As an
immediate consequence of this finding ($\tau_D \gg \tau_i$) we
predict, that the measured 2D electron (hole) mobility in each layer
would be completely independent of the carrier density in the {\it other}
layer. This prediction has indeed been explicitly verified
experimentally \cite{Seamons2}.

In conclusion, we theoretically study transport and frictional drag
of  undoped electron-hole bilayers based on HIGFETS
within a many-body Fermi liquid theory. 
We find that  the unintentional
background charged impurity scattering  
is the most dominant resistive scattering  mechanism in the recent experimental
systems \cite{Seamons,Seamons2}. This implies that the 2D mobility in
each layer is 
independent of the carrier density in the other layer in the
high-mobility structures of refs. \onlinecite{Seamons,Seamons2}.
We also find that the drag resistivity is 
significantly enhanced when the electron-hole layer separation is
small, 
but our Fermi liquid many-body approach cannot explain the recently
observed\cite{Seamons2} upturn in the drag resistance with the lowering of
temperature, which may be indicating the emergence of a non-Fermi
liquid excitonic phase in closely spaced bilayers.
Our Fermi liquid theory would always predict $\rho_D \rightarrow 0$ as
$T \rightarrow 0$, and $\rho_D$ increasing with decreasing $T$ is an
unexplained anomaly in the experimental observation of
Ref. \onlinecite{Seamons2}.

This work is supported by DOE through Sandia National Laboratories.

\end{document}